\documentclass[conference]{IEEEtran}
\IEEEoverridecommandlockouts
\usepackage{cite}
\usepackage{algorithmic}
\usepackage{graphicx}
\usepackage{textcomp}
\usepackage{xcolor}
\usepackage{float}
\usepackage{subfig}
\usepackage{hyperref}
\usepackage{caption}
\usepackage{mathptmx}  

\usepackage[T1]{fontenc}
\usepackage[utf8]{inputenc}

\def\BibTeX{{\rm B\kern-.05em{\sc i\kern-.025em b}\kern-.08em
    T\kern-.1667em\lower.7ex\hbox{E}\kern-.125emX}}

\begin{document}
\title{NeuroVM: Dynamic Neuromorphic Hardware Virtualization}

\author{%
\IEEEauthorblockN{Murat Isik}
\IEEEauthorblockA{%
Stanford University\\
Stanford, USA \\
Email: mrtisik@stanford.edu}
\and
\IEEEauthorblockN{Jonathan Naoukin}
\IEEEauthorblockA{%
University of Texas at Austin\\
Austin, USA \\
Email: jnaoukin@utexas.edu}
\and
\IEEEauthorblockN{I. Can Dikmen}
\IEEEauthorblockA{%
Temsa Research \& Development Center\\
Adana, Turkey \\
Email: can.dikmen@temsa.com}
}

\maketitle

\section*{Abstract}

This paper introduces a novel approach in neuromorphic computing, integrating heterogeneous hardware nodes into a unified, massively parallel architecture. Our system transcends traditional single-node constraints, harnessing the neural structure and functionality of the human brain to efficiently process complex tasks. We present an architecture that dynamically virtualizes neuromorphic resources, enabling adaptable allocation and reconfiguration for various applications. Our evaluation, using diverse applications and performance metrics, provides significant insights into the system's adaptability and efficiency. We observed scalable throughput increases across configurations of 1, 2, and 4 Virtual Machines (VMs), reaching up to 5.1 Gibibits per second (Gib/s) for different data transfer sizes. This scalability demonstrates the system's capacity to handle tasks that require substantial amounts of data. The energy consumption of our virtualized accelerator environment increased nearly linearly with the addition of more NeuroVM accelerators, ranging from 25 to 45 millijoules (mJ) as the number of accelerators increased from 1 to 20. Further, our investigation of reconfiguration overheads revealed that partial reconfigurations significantly reduce the time spent on reconfigurations compared to full reconfigurations, particularly when there are more virtual machines, as indicated by the logarithmic scale of time measurements.

\section{Introduction}

Over the past decade, the landscape of computation architectures has changed significantly, moving away from traditional CPU-based systems toward specialized, parallel architectures such as GPUs and FPGAs. The advancements have improved processing power, energy efficiency, and the ability to adapt to a variety of workloads. However, with the increasing complexity of computational tasks, particularly in artificial intelligence and machine learning, new paradigms for computing are needed\cite{bobda2022future, nurvitadhi2016accelerating, wyant2012computing, isik2023survey, vogginger2024neuromorphic}.

Neuromorphic computing, inspired by the structures and functions of the human brain, offers a transformational solution. Synaptic connections and neural networks are emulated by neuromorphic systems in order to facilitate complex cognitive tasks. In addition to their suitability for artificial intelligence, robotics, and scientific computations, these systems rely on parallel processing and adaptive learning principles\cite{isik2024advancing,isik2024accelerating}. Multi-node neuromorphic computing systems leverage the capabilities of neuromorphic computing by integrating heterogeneous hardware. A single node cannot address computations beyond its capabilities when multiple architectures are integrated\cite{qin2017, du2015neuromorphic, balaji2019framework, moreira2020neuronflow, isik2023design, fanuli2024allowing}. It is challenging to develop scalable and portable software technologies for large-scale neuromorphic systems as well as to explore virtualization within them. FPGAs have been rapidly adopted across a wide range of applications due to their versatility. The time-consuming nature of feature extraction algorithms makes them unsuitable for real-time applications. To address this issue, dedicated hardware is used, such as FPGAs, which allow complex operations to be performed in parallel. As a result of FPGA virtualization, the hardware is abstracted, the interface is decoupled, and the complexity of the framework is hidden\cite{vaishnav2018survey, quraishi2021survey, zeng2020enabling, yazdanshenas2017quantifying, wang2024low}. FPGA virtualization methodologies and definitions have evolved with changing application requirements. An innovative neuromorphic architecture leveraging dynamic virtualization is proposed in this paper to tackle these challenges. Through exploring the integration, virtualization, and optimization of neuromorphic hardware nodes, we aim to realize the full potential of neuromorphic computing for a wide range of scientific and computational applications. Our contributions include:

\begin{itemize}
\item Proposing a novel neuromorphic architecture that integrates multiple hardware nodes through dynamic virtualization, enhancing the system's ability to handle complex computations and adapt to varying workloads.
\item Conducting an extensive analysis of key performance indicators such as throughput, energy efficiency, and resource utilization, and performing a comparative study with traditional single-node neuromorphic systems to highlight the advantages of our multi-node, virtualized architecture.
\item Outlining future research directions, focusing on integrating specialized accelerators with the neuromorphic fabric and exploring security implications in virtualized neuromorphic environments.
\end{itemize}

This paper investigates the union of diverse neuromorphic hardware nodes within a parallel framework. \textbf{Section 2} reviews the history and concept of virtualization in neuromorphic computing. \textbf{Section 3} explores the architectural design and the dynamic virtualization essential for resource management. \textbf{Section 4} assesses the system's performance using different VM configurations. \textbf{Section 5} concludes by summarizing the study's primary insights. Finally, \textbf{Section 6} envisions future work on integrating specialized accelerators and securing virtualized neuromorphic environments.

\section{Background}

Neuromorphic computing, inspired by the human brain, represents a major advancement in artificial cognition. These systems mimic neural networks and synaptic connections, crucial for AI, robotics, and complex data analysis. However, single-node hardware configurations limit their potential. Transitioning to integrated, parallel systems is natural but challenging, especially regarding software compatibility and adaptability with expansive neuromorphic setups.

Our goal is to create a unified system architecture to efficiently manage tasks across neuromorphic hardware nodes, each contributing unique capabilities. The system is designed to be flexible and self-optimizing in response to varying workloads as other researchers set up a direction \cite{nilsson2023integration, huynh2022implementing}. We are exploring neuromorphic virtualization for dynamic reconfiguration and resource sharing, aiming to enhance the adaptability, efficiency, and performance of neuromorphic systems to move forward from their results \cite{telecom4040032, pfeiffer2018deep, isik2023astrocyte}.

Virtualization has become fundamental in computing, optimally allocating capabilities between hardware and OS. Conceived by IBM in the 1960s to partition mainframes into multiple virtual instances, virtualization has evolved to improve efficiency and reduce costs. The hypervisor, or virtual machine monitor (VMM), abstracts physical resources from the OS, enabling multiple OSs to run simultaneously on a single hardware platform \cite{schmid2020accessible, li2019time, brant2012zuma}. This enhances resource utilization and strengthens security, reliability, and resilience. With cloud computing's rise, virtualization bridges hardware and software applications, creating a cohesive operational ecosystem. Various virtualization techniques, including full, OS-layer, hardware, para, application, and resource virtualization, offer distinct benefits in resource sharing, isolation, and efficiency.

FPGA virtualization is not new; various microkernel-based approaches have been explored, enabling mapping and exchange of hardware accelerators to VMs. However, these systems focused mainly on resource utilization, neglecting energy efficiency and diverse guest OS requirements.

The FPGA virtualization layer L4ReC was developed to enable shared use of reconfigurable resources by multiple guest OS while addressing embedded systems' constraints. It tackles key challenges such as system isolation—ensuring performance and data isolation to prevent interference and security breaches. L4ReC integrates FPGA virtualization into the L4Re micro-hypervisor, creating virtual FPGAs (vFPGAs) that allow VMs to access hardware accelerators regardless of location. A crucial feature is the dynamic mapping and scheduling of hardware threads, prioritizing real-time needs. Early results on Xilinx Ultrascale MPSoC show improved FPGA resource utilization and energy efficiency, especially in battery-powered devices. L4ReC represents a significant step forward in optimizing reconfigurable resources in embedded systems \cite{wulf2022virtualization}.

Authors \cite{bandara2022enabling} explore Linux's foundational concepts and technologies underpinning the Virtio-FPGA solution. They provide background on the Virtio standard, the Linux FPGA Manager component of the Linux kernel, and the VFIO pass-through for ARM in QEMU and Device Tree Overlays technologies. FPGA overlay architectures on computational storage devices enable programmable near-storage processing. These architectures consist of reconfigurable operators, crossbar stream switches, and on-board DRAM, facilitating data movements between operators and the FPGA's DRAM. The storage interfaces in these architectures allow direct access to storage units, adopting the NVMe standard for fast and parallel access. Software support for these systems includes an abstraction layer simplifying near-storage processing, exposing FPGA overlay architecture operators as executable files within an OS. In I/O virtualization, software-based virtualization presents virtual device instances for device sharing across VMs. Full virtualization uses a trap-and-emulate approach, while paravirtualization creates VM-friendly virtual device interfaces.

\begin{table*}[h]
\centering
\caption{Comparison of Neuromorphic Virtualization vs. Digital Virtualization}
\label{tab:neuro_vs_digital_virt}
\begin{tabular}{|p{3.5cm}|p{4.5cm}|p{4.5cm}|p{4.5cm}|}
\hline
\textbf{Functional Hierarchy} & \textbf{Key Technology} & \textbf{Neuromorphic Virtualization Solution} & \textbf{Digital Virtualization Solution} \\ \hline
Resource Pool Management & Integrated neuromorphic resources, centralized management, dynamic allocation, monitoring, maintenance, and unified scheduling. & Use neuromorphic-specific management tools for resource pools. & Use traditional virtualization management tools like VMware, and Hyper-V. \\ \hline
Virtualization Layer & Neuromorphic virtualization, VM management, and container management. & Employ neuromorphic-specific virtualization technologies for resource abstraction. & Utilize hardware-assisted or software-assisted virtualization technologies. \\ \hline
Resource Isolation Layer & Hardware isolation technology, software isolation technology, or a combination of the two. & Implement neuromorphic-specific isolation technologies to prevent resource contention. & Use established isolation technologies like Intel VT-x, and AMD-V. \\ \hline
Scheduling Layer & Neuromorphic resource scheduling algorithm, load balancing, resource prediction. & Developing and use neuromorphic-aware scheduling algorithms for efficient resource allocation. & Apply conventional scheduling and load balancing algorithms. \\ \hline
Application Layer & Neuromorphic task scheduling algorithm, dynamic migration technology, application deployment, and management. & Utilize algorithms optimized for neuromorphic computing tasks and dynamic resource management. & Leverage general-purpose computing algorithms and migration technologies like live VM migration. \\ \hline
\end{tabular}
\end{table*}

\section{Methodology}

Neuromorphic hardware, evolving from basic silicon neurons to advanced neuromorphic chips, offers significant advantages in dynamic resource management and virtualization. The integration of FPGAs with neuromorphic systems brings new capabilities in virtualization and resource utilization. This section explores methodologies and implications of virtualizing neuromorphic hardware, focusing on design, application, and system architecture. Initial efforts aimed to replicate brain neural structures in silicon, progressing to sophisticated neuromorphic chips for various fields. FPGAs are ideal for neuromorphic systems due to their flexibility and reconfigurability, aligning with efficient computing models. Virtualization enables dynamic resource management, meeting variable computational demands by reconfiguring resources. Studies have explored design methodologies for dynamic management and reconfiguration of neuromorphic systems \cite{nilsson2023integration, johnson2022hardening, pachideh2023towards, kumar2023digital}.

Neuromorphic hardware, designed to emulate biological neurons, faces growing computational demands, especially for Spiking Neural Network (SNN) inference. Neuromorphic Hardware Virtualization tackles these challenges through dynamic resource allocation and reconfiguration, optimizing performance while minimizing power and hardware use. Neurons communicate via spikes, reducing logic load, which is essential for NeuroVM. This framework combines hardware design, security, and neuromorphic computing to create adaptable and secure systems for various tasks.

\begin{figure}[h]
\centering
\includegraphics[scale=0.45]{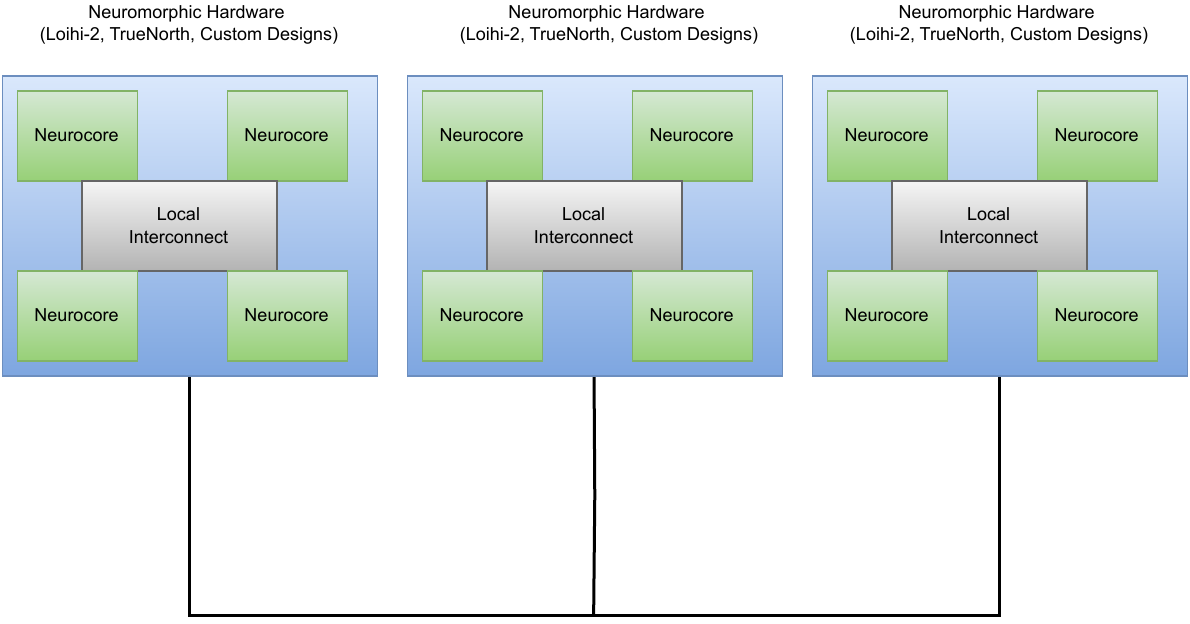}
\caption{Schematic of Neurocore Interconnectivity in Neuromorphic Hardware}
\label{fig1}
\end{figure}

\begin{figure}[h]
\centering
\includegraphics[scale=0.5]{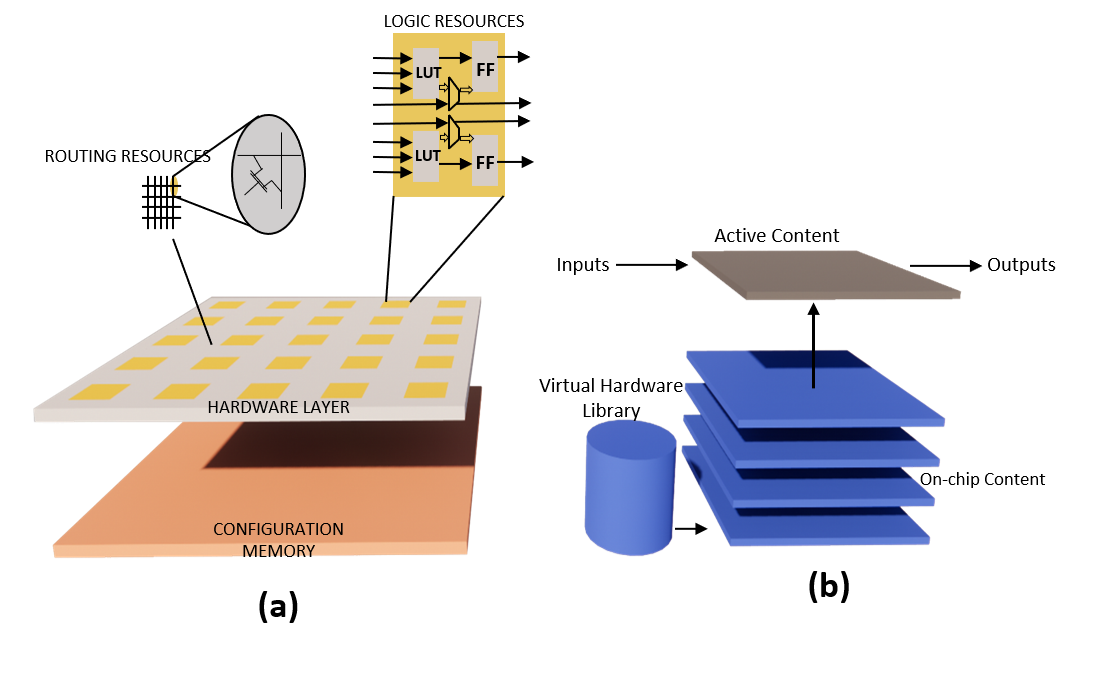}
\caption{FPGA Virtualization Concept}
\label{fig12}
\end{figure}

The virtualization of neuromorphic hardware is key to our research, offering greater efficiency and flexibility. By optimizing task profiling, memory, and interconnects, along with a sophisticated kernel controller, we aim to advance neuromorphic computing. As shown in \autoref{fig1}, neurocores—modular processing units that emulate brain circuits are interconnected to enable parallel processing, which is crucial for high-speed, energy-efficient computations in neural network simulations.Neuromorphic computing necessitates hardware platforms capable of emulating neural architectures. FPGAs are well-suited for this due to their reconfigurable nature and parallel processing capabilities. FPGA virtualization abstracts physical FPGA resources to enable multiple applications on a single chip. This section details methodologies and strategies in FPGA virtualization, emphasizing Dynamic Function Exchange (DFX) for adaptability and efficiency, conceptually illustrated in \autoref{fig12}. DFX is an FPGA feature facilitating runtime reconfiguration of hardware functions without disrupting system operation. DFX allows selective activation and deactivation of hardware modules, enabling the FPGA to adapt to new tasks on-the-fly. This is crucial in neuromorphic computing, where dynamic reconfiguration in response to neural network demands is essential.

Our approach leverages DFX to create a virtualized neuromorphic hardware environment, enabling real-time reconfiguration of neural network models. A dedicated controller manages the dynamic loading and unloading of DFX modules, partitioning hardware resources for concurrent tasks while ensuring data isolation and security. Virtualization improves adaptability and performance through resource sharing. For I/O, software mechanisms such as full and paravirtualization facilitate device sharing across multiple VMs, with paravirtualization reducing overhead. Our methodology includes an advanced task profiling system that assesses tasks based on processing demands and parallel execution suitability. This guides task allocation to appropriate neuromorphic nodes, optimizing resource utilization and computational throughput. We design specialized memory hierarchies and interconnect architectures tailored to neuromorphic hardware, aiming to establish a high-bandwidth, low-latency communication framework for efficient data exchange across the neuromorphic network. The kernel controller driver manages data influx and controls neuromorphic processor cores. Inspired by existing paradigms like the Coyote driver, our design virtualizes neuromorphic processors, replacing traditional VFPGAs with bespoke designs. The driver interfaces with user space through C++ constructs, ensuring seamless hardware interaction.

FPGA virtualization segments FPGA capabilities to mimic multiple discrete processing units, crucial for resource management similar to virtual machines. Our project uses the Vitis platform for FPGA development, leveraging C++ and embedded systems expertise to implement PS-PL relationships within the architecture. We focus on high-level synthesis in Vitis to meet neuromorphic virtualization needs. Table \ref{table:utilization} summarizes the resource utilization for the Zynq UltraScale+ XCZU7EV MPSoC, detailing used and available resources such as Logic Cells, Memory, DSP Slices, and I/O Pins, along with utilization percentages for reference. Our system addresses resource contention by employing a dynamic scheduling algorithm that prioritizes tasks based on real-time requirements and resource availability. Each VM operates within a defined allocation of neuromorphic resources, and when contention occurs, the system dynamically reallocates resources to ensure tasks maintain performance. Additionally, the architecture incorporates performance isolation mechanisms to reduce the impact of contention on overall system efficiency. 

\begin{table}[h]
\centering
\caption{Resource utilization summary}
\label{table:utilization}
\begin{tabular}{|c|c|c|c|}
\hline
\multicolumn{4}{|c|}{Zynq UltraScale+ XCZU7EV} \\ \hline
\textbf{Resource} & \textbf{Utilization} & \textbf{Available} & \textbf{\% Utilization} \\ \hline
LUT & 151,200 & 504,000 & 30 \\
Memory & 11.4MB & 38MB & 30 \\
IO & 139 & 464 & 29.19 \\
DSP & 518 & 1,728 & 29.94 \\ \hline
\end{tabular}
\end{table}

\begin{figure*}[h]
\centering
\includegraphics[scale=0.45]{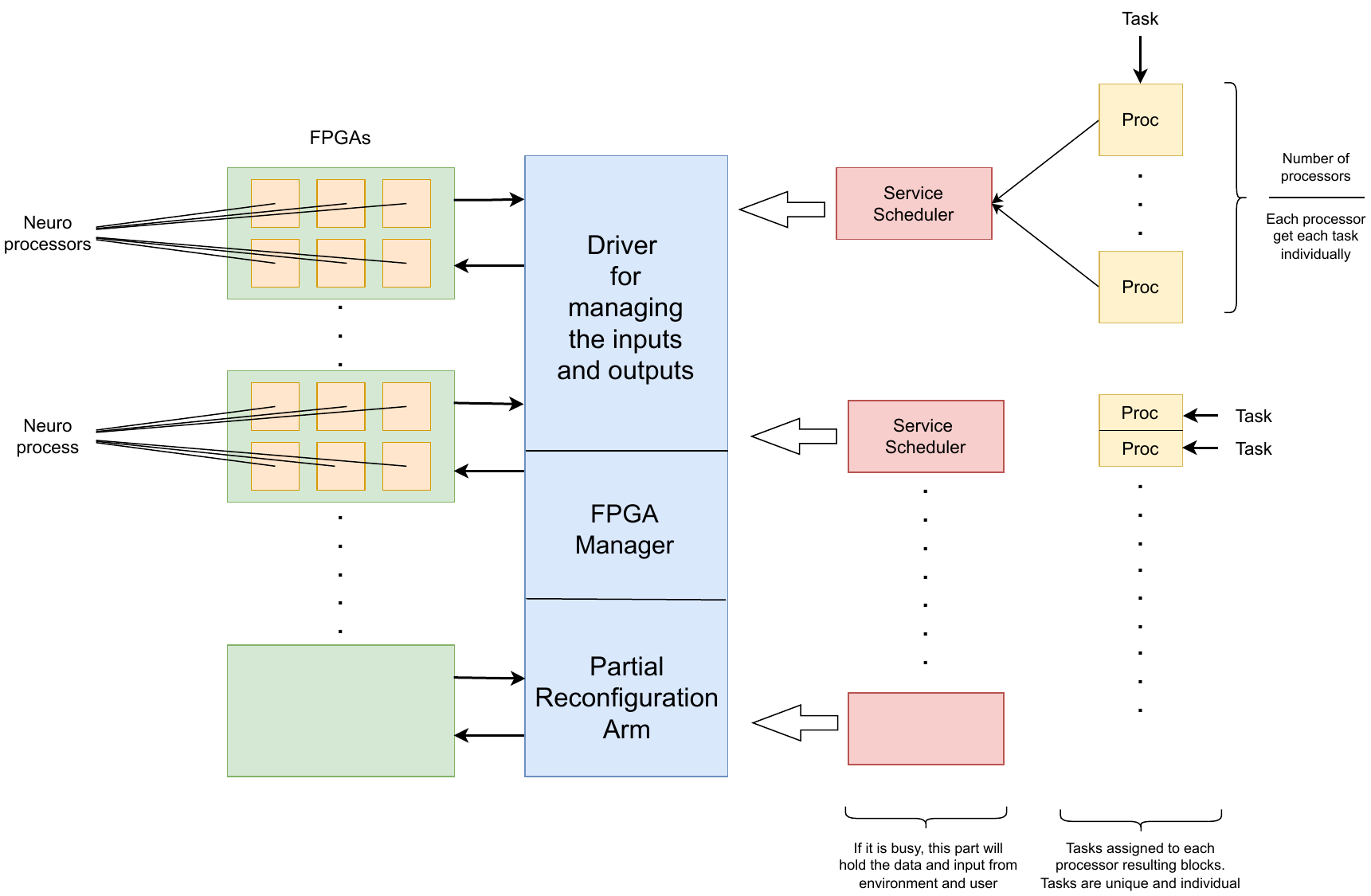}
\caption{Task Management and Processing Architecture in Neuromorphic Systems}
\label{fig6}
\end{figure*}

\autoref{fig6} shows the task management and processing architecture in a neuromorphic computing system. The 'Driver' handles system inputs and outputs, interfacing with internal processing units. Below it, the 'Service Scheduler' efficiently allocates tasks to 'Neuroprocessors' optimized for neural network simulations. The 'Proc' blocks represent individual processing units executing tasks, while the 'FPGA Manager' oversees on-the-fly hardware reconfiguration. The system includes buffering to manage peak processing times and ensures smooth task execution across multiple dedicated hardware units. In our system, communication between nodes is facilitated by a high-bandwidth, low-latency interconnect architecture specifically designed for neuromorphic hardware. We mitigate potential communication bottlenecks by optimizing task profiling and dynamic resource allocation.  Parallel processing is ensured by the system's modular neurocores, which are interconnected to ensure efficient data exchange. Additionally, we incorporate advanced scheduling strategies to minimize data transfer delays, especially when handling intensive, real-time tasks. As a result, our architecture supports scalability and performance in distributed, multi-node environments.

\section{Evaluation}

The key performance indicators we measured were throughput, energy efficiency, resource utilization, and reconfiguration overhead. Scalability, adaptability, and energy consumption are among the architecture's strengths, demonstrating its advantages over previous designs. 

\begin{figure}[h]%
    \centering
    \begin{minipage}{0.48\textwidth}
        \centering
        \subfloat[Throughput\label{fig:throughput_transfer_size}]{{\includegraphics[width=\linewidth]{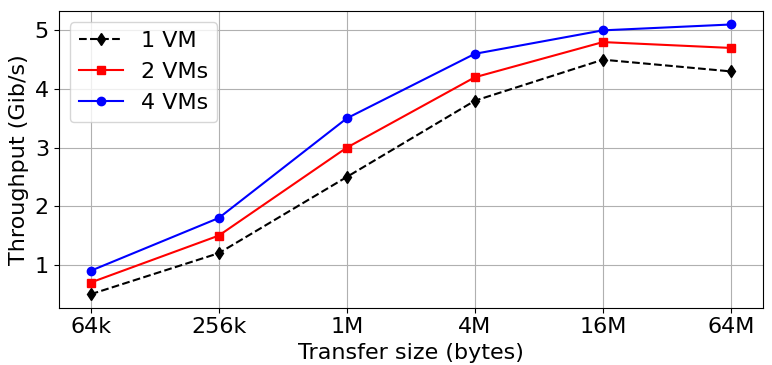} }}
    \end{minipage}%
    \hfill 
    \begin{minipage}{0.48\textwidth}
        \centering
        \subfloat[Hardware Resource Utilization\label{fig:resource_utilization}]{{\includegraphics[width=\linewidth]{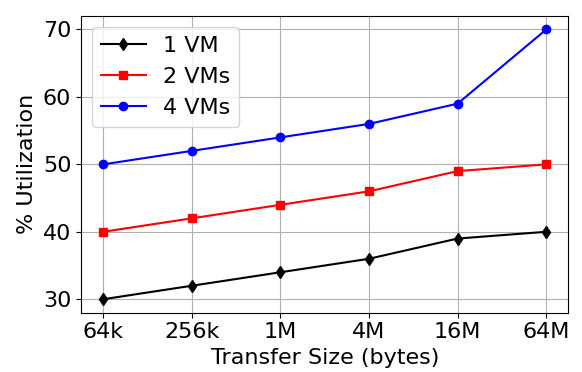} }}
    \end{minipage}
    \caption{Comparative analysis of NeuroVM performance.}
    \label{fig:vm_performance_analysis}
\end{figure}

\begin{figure}[h]
\centering
\includegraphics[scale=0.7]{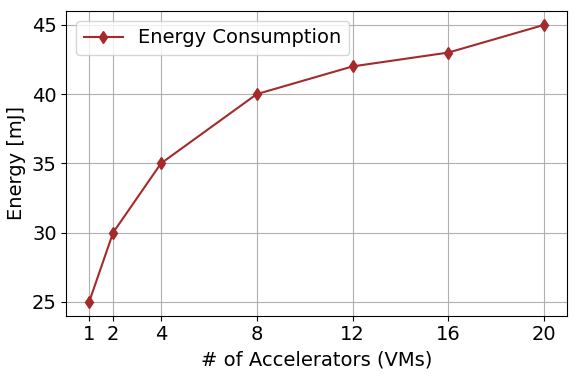}
\caption{NeuroVM Energy Consumption.}
\label{fig9}
\end{figure}

\autoref{fig:vm_performance_analysis} (a) illustrates the relationship between transfer size and throughput in Gibibits per second (Gib/s) for 1, 2, and 4 VMs. VM configurations are represented by different colors and markers, revealing how throughput scales as transfer sizes increase. These three lines provide insight into the efficiency and scalability of the VMs under varying data transfer loads, with the black dashed line representing one VM, the red solid line representing two VMs, and the blue dotted line representing four VMs. This analysis is performed to understand the performance dynamics for neuromorphic computing designs. \autoref{fig:vm_performance_analysis} (b) depicts the percentage of resource utilization against different transfer sizes for 1, 2, and 4 VMs.  VM utilization efficiency is shown by different colored lines and markers. This graph shows the results for neuromorphic VM efficiency by showing how resource utilization varies with throughput size. The utilization increases with larger throughput sizes, especially in more VM-based configurations, highlighting the importance of VM density in neuromorphic computing environments.  \autoref{fig9} illustrates the energy consumption trends in a virtualized accelerator environment.  According to the results, as the number of NeuroVM accelerators increases, the energy demand also increases. Understanding this trend is of great importance, especially in the context of neuromorphic computing, where energy management is an important performance metric. The almost linear increase in energy consumption with more accelerators reveals the need to optimize resource allocation to balance computational power and energy efficiency.

\begin{figure}[h]
\centering
\includegraphics[scale=0.58]{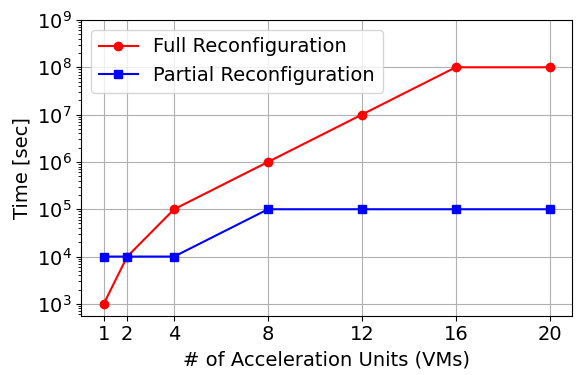}
\caption{NeuroVM implementation overhead with full and partial reconfigurations.}
\label{fig10}
\end{figure}

\autoref{fig10} presents a comparative analysis of the implementation overhead for full and partial reconfigurations in a NeuroVM environment. Full reconfigurations are shown with a red line with circle markers, while partial reconfigurations are shown with a blue line with square markers. In particular, partial reconfiguration shows a significantly reduced time overhead compared to full reconfiguration, especially as the number of virtual machines increases. In neuromorphic virtual machines, where reconfiguration times play a critical role in overall system efficiency and responsiveness, this observation is crucial for optimizing performance. Partial reconfiguration strategies can benefit dynamic computing environments that require rapid adaptation. Through this comprehensive evaluation, we aimed to demonstrate the robustness, versatility, and efficiency of our neuromorphic computing architecture. Our design is designed to address a variety of applications and leverage the benefits of virtualization, and showcases the potential of virtualized neuromorphic hardware in shared computing situations.

\section{Conclusion}

We evaluated an innovative neuromorphic computing architecture that demonstrates the potential of high-performance computing to be a unique field. Throughput, energy efficiency, resource utilization, and reconfiguration overhead are tested for our multi-node, virtualized neuromorphic architecture. According to the accompanying figures, this architecture offers significant advantages over traditional single-node neuromorphic systems. Partially reconfiguring virtual machines significantly reduces the overhead compared to a full reconfiguration, especially as the number of virtual machines increases. As a result of this approach, computational interruptions are minimized and tasks with tight deadlines are not delayed by reconfigurations. It will be important to compare our system with other multi-node neuromorphic architectures.

\section{Future Works}
We plan to compare our approach with multi-node neuromorphic systems to determine its advantages and limitations. This work will include benchmarking performance metrics such as throughput, latency, power consumption, and reconfiguration overheads on various systems. Our future efforts will focus on several goals to improve our proposed neuromorphic architecture. The primary goal will be the integration of specialized accelerators to increase computational efficiency and expand the ability to handle complex, intensive tasks such as advanced machine learning and real-time analytics. Additionally, we plan to address security challenges in virtualized neuromorphic environments by developing protocols to mitigate threats and ensure system integrity. We will continue to optimize the architecture to meet the specific needs of a variety of applications, from scientific simulations to industrial processing.


\bibliographystyle{IEEEtran}

\bibliography{external}

\end{document}